\documentclass[twocolumn,superscriptaddress]{revtex4}
\usepackage[dvips]{graphicx}
\usepackage{times}
\usepackage{amsmath}
\begin{document}
\title{%
Notes on the bunching peak of $g^{(2)}$ functions for correlated photons from 
single quantum dots
}
\author{T.~Kuroda}
\affiliation{%
National Institute for Materials Science, 
1-1 Namiki, Tsukuba 305-0044, Japan}
\affiliation{%
Universit\`{a} di Firenze, via G. Sasone 1, Sesto Fiorentino 50019, Italy}
\author{C.~Mastransdrea}
\author{M.~Abbarchi}
\author{M.~Gurioli}
\affiliation{%
Universit\`{a} di Firenze, via G. Sasone 1, Sesto Fiorentino 50019, Italy}
\date{\today}
\begin{abstract}
Second-order correlation functions for photon pulses associated with exciton-biexciton cascades are theoretically derived. A finite efficiency in photon detection and statistical distribution in exciton numbers are taken into account. It is found that the bunching peak height of photon statistics ($g^{(2)}(0)$) depends on the mean number of excitons, $\bar{N}$, and significant bunching is only detectable at very low excitation, $\bar{N}\ll 2$. 
\end{abstract}
\maketitle
Semiconductor quantum dots (QDs) are promising objects which produce nonclassical photons. Since the first demonstration of photon antibunching in 2000, many attempts have been done for developing efficient single-photon sources. Another hot topic in this research area is the generation of entangled photon pairs exploiting biexciton--exciton cascades. These experiments commonly utilize a Humbury-Brown and Twiss (HBT) setup, measuring coincidence photon counts. In this note we derive a simple form for a histogram of coincidence, representing second-order correlation functions, $g^{(2)}(\tau)$,  for correlated photons associated with biexciton--exciton cascades after pulsed excitation. 

\begin{figure}
\includegraphics[scale=0.6]{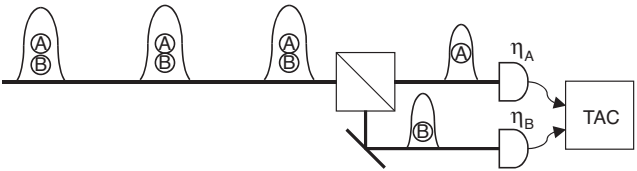}
\caption{%
A typical HBT setup that measures $g^{(2)}$ functions for correlated photon pulses, each of which contains photon A and photon B.}
\label{fig_1}
\end{figure}

A typical HBT setup is illustrated in Fig.~\ref{fig_1}. Photon pulses emitted from a single QD are divided by a beam splitter. Each photon (photon A and B) is detected by a pair of photon-counting detectors, whose yields are $\eta_A$ and $\eta_B$. Electrical outputs from the detectors act as start and stop pulses for a time-to-amplitude converter (TAC), and measuring coincidence events as a function of time delay. For pulsed excitation, a histogram generally exhibits a central peak following sequential peaks with an interval of laser repetition, $T_{rep}$.  Here we assume that histograms at positive delays correspond to events when photon A (B) starts (stops) TAC, while those at negative delays show the events of reversed order.

First we consider a very simple situation, where each pulse contains photon A and photon B. This is the case when correlated photons are perfectly bunched and regulated -- see Fig.~\ref{fig_1}. In this case, coincidence at $\tau=0$ is simply proportional to the product of detector yields, 
\begin{equation}
g^{(2)}(0) = \eta_A \times \eta_B.
\end{equation}

Note that for ideal detectors with $\eta_A=\eta_B=1$, the function of $g^{(2)}(\tau)$ shows unity at $\tau=0$ with absence of side peaks at $|\tau|>0$, exhibiting a perfect bunching feature. However, for realistic detectors with finite yields, there is a probability of missing a stop photon, and waiting for a delayed pulse to stop TAC. This leads to a decrease in a central peak, and emergence in side peaks in histograms. For example, coincidence at $\tau=T_{rep}$ is given by the product of (i) probability for detector A to start TAC, (ii) probability for detector B to be failed to count at the same time, $\tau=0$, and (iii) probability for detector B to count at $\tau=T_{rep}$ and stop TAC. Thus, we get
\begin{equation}
g^{(2)}(T_{rep}) = \eta_A (1-\eta_B) \eta_B.
\end{equation}
Following the same manner, we find the intensity of the $n$-th ($m$-th) side peak, 
\begin{equation}
\begin{split}
g^{(2)}(+nT_{rep}) &= \eta_A (1-\eta_B)^n \eta_B \\
&= g^{(2)}(0)(1-\eta_B)^n,
\end{split}
\end{equation}

\begin{equation}
\begin{split}
g^{(2)}(-mT_{rep}) &= \eta_B (1-\eta_A)^m \eta_A \\
&= g^{(2)}(0)(1-\eta_B)^n, 
\end{split}
\end{equation}

\begin{figure}
\includegraphics[scale=0.75]{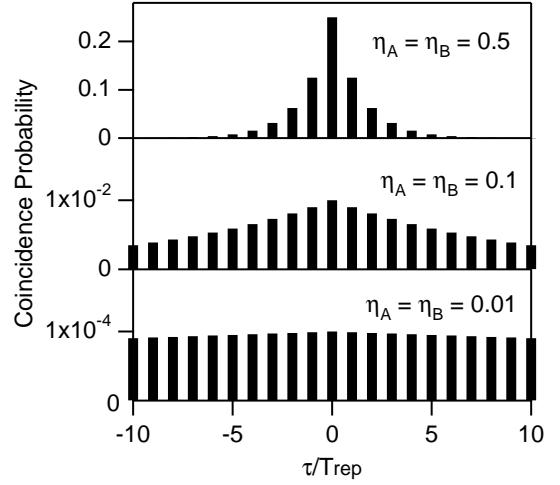}
\caption{%
Coincidence probability for perfectly regulated photon-pair pulses using detectors with various yields, $\eta$.}
\label{fig_2}
\end{figure}

The relevant histogram is illustrated in Fig. 2. It shows a central peak at $\tau=0$ following multiple side peaks whose envelope decays with $(1-\eta)$. We should note that the value of $\eta$ may decrease to $<10^{-2}$, since the detection yield in our treatment characterizes every lossy processes, such as photon extraction from QDs, mode coupling, and photon counting at detectors. This leads to a striking result: a bunching peak becomes quite small for regulated pulses each of which contains two photons. 

Next we will take into account distribution in the number of photons. This is the case when photon pulses  are not regulated. We assume that the number of photons is determined by the number of excitons, obeying a Poisson distribution, 
\begin{equation}
P_{\bar{N}}(n)=\exp(-\bar{N})\frac{\bar{N}^n}{n!}, 
\label{poisson}
\end{equation}
where $\bar{N}$ is the mean exciton number, which is proportional to excitation density. Here we detect photons with a biexciton energy ($B$) and those of a exciton energy ($X$) emitted from a single QD. A photon $X$ is generated when the number of excitons is more than one, and both photons $B$ and $X$ are generated when more than two excitons are present initially. Thus, probability to find photons $X$ and $B$ in a pulse, $P_X$ and $P_B$, respectively, is; 
\begin{align}
P_X &= \sum_{n\ge1}P(n)\notag\\
&=1-P(0)\notag\\
&=1-\exp(-\bar{N}),\\
P_B &= \sum_{n\ge2}P(n)\notag\\
&=1-P(0)-P(1)\notag\\
&=1-\exp(-\bar{N})(1-\bar{N}).
\end{align}
where we use a relation, $\sum_{n\ge0}P(n)=1$. Note that if the mean number of excitons is sufficiently large so that $\bar{N}\gg2$, it is highly probable for both $B$ and $X$ photons to be emitted, approaching the case of regulated photon pulses discussed above. 

Coincidence at $\tau=0$ is given by a joint probability for both $B$ and $X$ photons to be present in a pulse, and to be counted by two detectors, $B$ and $X$. Thus we find, 
\begin{equation}
\begin{split}
g^{(2)}(0) &=  \sum_{n\ge2}P(n)\eta_B\eta_X \\
&=P_B\eta_B\eta_X.
\end{split}
\end{equation}

Coincidence at $\tau=T_{rep}$, the first side peak in histogram, is given by the product of (i) probability for photon $B$ to be present in a pulse, (ii) probability for detector $B$ to start TAC, (iii) probability for detector $X$ to be failed to count, (iv) probability for photon $X$ to be present at a pulse delayed by $T_{rep}$, and (v) probability for detector X to stop TAC at $\tau=T_{rep}$. Thus, we find, 
\begin{equation}
\begin{split}
g^{(2)}(T_{rep}) &=P_B\eta_B(1-\eta_X)P_X\eta_X \\
&=g^{(2)}(0)P_X(1-\eta_X).
\end{split}
\end{equation}
Following the same manner, we find expression for the $n$-th side peak at positive delays, 
\begin{equation}
\begin{split}
g^{(2)}(nT_{rep}) &= P_B\eta_B(1-\eta_X) \\
&\times [P(0)+P_X(1-\eta_X)]^{n-1} P_X \eta_X \\
&=g^{(2)}(T_{rep})\{1-P_X\eta_X\}^{n-1} \\
&=g^{(2)}(0)P_X(1-\eta_X)\{1-P_X\eta_X\}^{n-1}.
\end{split}
\end{equation}

For negative delays we will consider the case when detector X (B) starts (stops) TAC.  We have that $g^{(2)}(-0)=0$ because the two photons appear due to relaxation cascades. For the first side peak at negative delays, we find
\begin{equation}
\begin{split}
g^{(2)}(-T_{rep}) &=[P(1)\eta_X+P_B\eta_X(1-\eta_B)] P_B\eta_B \\
&= g^{(2)}(0)(P_X-P_B\eta_B),
\end{split}
\end{equation}
and for the $m$-th side peak, we obtain, 
\begin{equation}
\begin{split}
g^{(2)}(-mT_{rep}) &= g^{(2)}(-T_{rep}) (1-P_B \eta_B)^{m-1} \\
&= g^{(2)}(0)(P_X-P_B\eta_B)(1-P_B \eta_B)^{m-1}.
\end{split}
\end{equation}

As we mentioned above, yield of photon detection is quite small so that $\eta\ll1$. In this case we obtain a simple expression, 
\begin{align}
g^{(2)}(\pm nT_{rep}) &\simeq g^{(2)}(0)P_X \notag\\
&= g^{(2)}(0)\{1-\exp(-\bar{N})\} \\
&=P_XP_B\eta_X\eta_B.
\end{align}
Equation 14 suggests the side peaks of histograms are identical to coincidence probability for uncorrelated photon pairs. 

\begin{figure}
\includegraphics[scale=0.75]{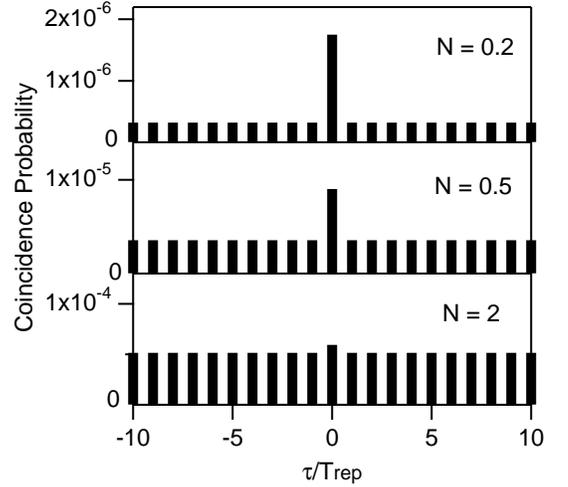}
\caption{%
Coincidence probability for correlated photon-pair pulses for various values of mean exciton numbers, $N$, with $\eta_B=\eta_X=10^{-2}$.}
\label{fig_3}
\end{figure}

Figure~\ref{fig_3} illustrates the relevant histogram for various values of $\bar{N}$. It shows a central bunching peak following constant side peaks. The relative height of the central peak is found to depend on $\bar{N}$, and it is larger for a smaller number of $\bar{N}$. The bunching peak height is, therefore, expressed by, 
\begin{align}
g^{(2)}(0)/g^{(2)}(T_{rep})&= P_X^{-1}\\
&= \{1-\exp(-\bar{N})\}^{-1}.
\end{align}

\begin{figure}
\includegraphics[scale=0.7]{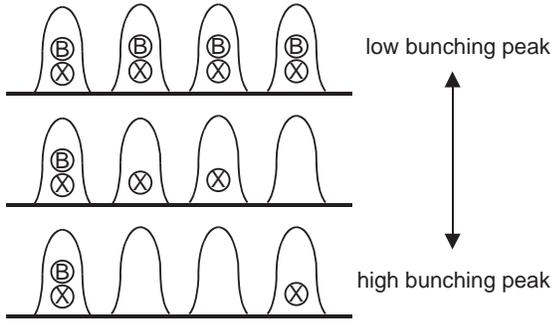}
\caption{%
Physical interpretation on the presence of a high bunching peak in $g^{(2)}$ functions.
}
\label{fig_4}
\end{figure}

Equation 15 suggests that a bunching peak is higher for larger probability to find pulses from "zero" excitons, reducing accidental coincidence counts. The physical situation is schematically illustrated by Fig.~4. 

\begin{figure}[t]
\includegraphics[scale=0.7]{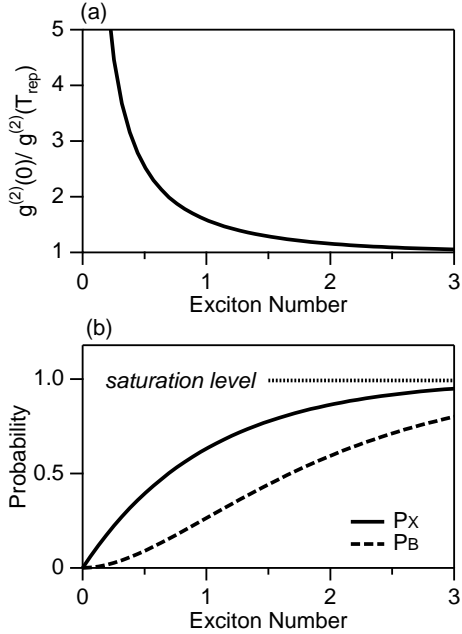}
\caption{%
(a) Relative height of bunching peaks, $g^{(2)}(0)/g^{(2)}(T_{rep})$, as a function of a mean number of excitons. For comparison, the same dependence of the intensity of exciton and biexciton fluorescence is plotted in (b).
}
\label{fig_5}
\end{figure}
 
Figure 5(a) plots the dependence of a bunching peak height on the mean number of excitons. As a comparison, the same dependence of probability to find photon $B$ and photon $X$ is presented by Fig.~5(b), which are proportional to the relevant counting rates. They show that to observe a significant bunching peak, a very small number of excitons, typically, less than 0.5, are required, although the X and B emissions intensities are much below their saturation levels. 

In conclusion, we derive a simple form of $g^{(2)}(\tau)$ for biexciton--exciton cascades. We found that a bunching peak should only appear at very low excitation when $\bar{N}\ll2$. This is a great contrast to an experimental condition for characterizing single-photon emitters, where one normally measures an anti-bunching dip at sufficiently high excitation, realizing regulated single-photon pulses. On the other hand, usage of \textit{dilute} photon pulses is essential to characterize a bunching feature for correlated photons.

\end{document}